\documentclass[prd,nofootinbib,12pt]{revtex4}

\usepackage{lineno,hyperref}
\usepackage{amsfonts,amssymb,amsmath}
\usepackage{epsfig}
\usepackage{mathrsfs}
\usepackage{amssymb}
\usepackage{graphicx}
\usepackage{amsmath}
\usepackage{graphicx}

\begin{document}

\title{Inflation in an effective gravitational model $\&$  asymptotic safety}
\author{Lei-Hua Liu$^{1}$}
\email{L.Liu1@uu.nl}
\author{Tomislav Prokopec$^{1}$}
\email{t.prokopec@uu.nl}
\author {Alexei A. Starobinsky$^{2,3}$}
\email{alstar@landau.ac.ru}
\affiliation{1. Institute for Theoretical Physics, Spinoza Institute and
the Center for Extreme Matter and Emergent Phenomena (EMME$\Phi$), Utrecht University,
Buys Ballot Building,
Princetonplein 5,
3584 CC Utrecht,
the Netherlands}
\affiliation{2. L.~D.~Landau Institute for Theoretical Physics RAS, Moscow 117334, Russian Federation}
\affiliation{3. Kazan Federal University, Kazan 420008, Republic of Tatarstan, Russian Federation}

\begin{abstract}

We consider an inflationary model motivated by quantum effects of gravitational and matter fields near the Planck 
scale. Our Lagrangian is a re-summed version of the effective Lagrangian recently obtained
by Demmel, Saueressig and Zanusso~\cite{Demmel:2015oqa} in the context
of gravity as an asymptotically safe theory. It represents a refined Starobinsky model,
${\cal L}_{\rm eff}=M_{\rm P}^2 R/2 + (a/2)R^2/[1+b\ln(R/\mu^2)]$,
where $R$ is the Ricci scalar, $a$ and $b$ are constants and $\mu$ is an energy scale.
By implementing the COBE normalisation and the Planck constraint on the scalar spectrum, we show that increasing 
$b$ leads to an increased value of both the scalar spectral index $n_s$ and the tensor-to-scalar ratio $r$. Requiring $n_s$ to 
be consistent with the Planck collaboration upper limit, we find that $r$ can be as large as $r\simeq 0.01$, the value 
possibly measurable by Stage IV CMB ground experiments and certainly from future dedicated space missions.
The predicted running of the scalar spectral index $\alpha=d n_s/d\ln(k)$ is still of the order $-5\times 10^{-4}$ (as 
in the Starobinsky model), about one order of magnitude smaller than the current observational bound.

\end{abstract}

\maketitle


\section{Introduction}

The $\Lambda CDM$ model supplemented with inflation is currently the best paradigm
that provides a consistent quantitative description for the accelerating expansion of universe,
cold dark non-baryonic matter (CDM), the origin of large scale structure (LSS) and temperature fluctuations in the 
cosmic microwave background (CMB). In addition, cosmic 
inflation~\cite{Starobinsky:1979ty,Starobinsky:1980te,Guth:1980zm,Linde:1981mu} provides the elegant and minimal 
solution of the horizon, flatness and homogeneity problems and can dilute magnetic monopoles if they were produced 
before inflation. In the most of inflationary models a scalar field -- dubbed {\it inflaton} -- is invoked to drive inflation 
and to seed primordial inhomogeneous adiabatic scalar perturbations  measured by observations at present. Inflaton can 
be a new scalar field that couples weakly to the Standard Model fields. (Such a coupling -- be it direct or indirect -- is 
required for successful post-inflationary stage of the energy transfer from inflaton to other quantum fields, (pre-)heating 
of matter and radiation and transition to the standard hot Big Bang.) Or it can be embedded into a (grand-unified) extension 
of the standard model, such as a Higgs-like field required for mass generation.

A notable alternative are purely geometric effective scalar degrees of freedom -- in this work we shall call these fields 
{\it scalarons} -- which can be generated by {\it e.g.} quantum-gravitational  effects. Namely, generically quantum fluctuations
of matter and gravitational fields at the Planck scale generate higher derivative local gravitational operators in the effective
action. If these operators are of the form of a function of the Ricci scalar $R$, $f(R)$,
by the method of Lagrange multipliers one can introduce a {\it scalaron} field which can play the role of inflaton.
Donoghue~\cite{Donoghue:1994dn}
pointed out that it is natural to expect that at energies much below the Planck scale only one of those higher dimensional operators
plays a significant role for the Universe's dynamics. The simplest such model was constructed already in 1980
by Starobinsky~\cite{Starobinsky:1980te}, and its predictions agree very well with current observations~\cite{Ade:2015lrj}.
The Lagrangian density in this model is of the form,
${\cal L} =(M_{\rm P}^2/2) R + a R^2/2$ where $a\approx 10^9\gg 1$, up to small one-loop corrections from matter
quantum fields which are responsible, in particular, for the scalaron decay and creation of all standard matter after the end of inflation,
see~\cite{Starobinsky:1981vz} for a more detailed quantitative description of the latter processes. 
The latter corrections which are of the type ${\cal R}\!\!{\cal R}\ln( {\cal R}\!\!{\cal R})$ at large curvatures, 
where ${\cal R}\!\!{\cal R}$ denotes some scalar quadratic 
combination built from the Riemann tensor, follow from perturbative quantum gravity~\cite{Stanyukovich:1965,DeWitt:1967yk,DeWitt:1967uc}, 
from the effective field theory approach to quantum gravity (conserving the local Lorentz invariance and general 
covariance)~\cite{Ginzburg:1971}, from the calculation of a renormalized average value of the energy-momentum tensor 
of quantum matter fields in external gravitational fields~\cite{Zeldovich:1971mw}, as well as by modern developments in this area.
However, once one-loop corrections have been taken into account, it is natural to think about higher-loop ones and try to account for 
them in some approximation. That is why in this paper we considered a refined Starobinsky model which is based on some assumption
about the form of re-summed logarithmic multi-loop corrections to the $R^2$ term.

While it is well known that General Relativity theory is non-renormalizable~\cite{'tHooft:1974bx,Goroff:1985sz,vandeVen:1991gw}, 
this is not so for the so called fourth-order gravity which contains terms $R^2$ and $W_{\alpha\beta\gamma\delta}W^{\alpha\beta\gamma\delta}$, 
where $W_{\alpha\beta\gamma\delta}$ is the Weyl tensor, in the Lagrangian density in addition to the Einstein term $R$. The fact that 
coefficients in front of these new terms are dimensionless (in particular, the $a$ coefficient in front of the $R^2$ term) already
suggests that this theory may be renormalizable in some sense, {\it e.g.} by power counting. However, it has a ghost in the tensor sector 
(though not in the scalar sector). Weinberg has proposed that gravity may be renormalizable in a weaker sense, which goes under the 
name of {\it asymptotic safety}~\cite{Weinberg:1976xy,Hawking:1979ig,Weinberg:2009bg,Weinberg:2016kyd}.
If gravity is an asymptotically safe theory, its ultraviolet completion would be given by a finite number of relevant operators,
and the corresponding coupling constants could be (in principle) determined by a finite number of measurements,
thereby making it predictable. Initially Weinberg proposed~\cite{Hawking:1979ig} that one could use renormalization group
methods in a small $\epsilon$ expansion around $D=2$, {\it i.e.} $D=2+\epsilon$, around $D=4$, but soon recognized that
that is not a very good expansion near $\epsilon\simeq 2$. A modern approach to study the ultraviolet sector of quantum gravity
uses functional renormalization group methods~\cite{Reuter:1996cp} (for a review see~\cite{Reuter:2012id})
and the results support the asymptotic safety hypothesis.
Recently, by working within the framework of asymptotic safety, Demmel, Saueressig and Zanusso~\cite{Demmel:2015oqa} have
assumed that quantum gravity in the ultraviolet may be represented by a series of local operators, starting
with $R$, $R^2$, {\it etc.} and
they have shown that the coefficient of $R^2$ runs approximately logarithmically with $R$.
This then implies that -- after resummation under some suitable assumptions-- at low energies the effective theory Lagrangian can
 be represented by
\begin{equation}
{\cal L} \simeq\frac{M_{\rm P}^2}{2}R+\frac{a R^2}{2[1+b \ln(R/\mu^2)]}
\,,
\label{EFT Lagrangian}
\end{equation}
where  $\mu$ is a renormalization scale and $a=a(\mu)\gg 1$ and $b=b(\mu)$ are $\mu-$dependent constants; {\it i.e.} once $\mu$ 
is fixed, the values of $a$ and $b$ are also fixed. Since the Hubble (curvature) scale during the observable part of inflation (last 
$50-60$ e-foldings)
is much below the Planck scale, $H_I\sim M_P/\sqrt{a} \sim 10^{14}~{\rm GeV}\ll E_{\rm P}\simeq 1.2\times 10^{19}~{\rm GeV}$,
it is reasonable to expect that the model Lagrangian~(\ref{EFT Lagrangian}) represents reasonably well the true effective theory of
modified gravity at inflationary curvatures. Furthermore, we expect that predictions of the inflationary model driven by the effective 
Lagrangian~(\ref{EFT Lagrangian}) do not differ by much from the predictions of the Starobinsky model. The analysis conducted in 
this work confirms that expectation, and moreover we provide an accurate answer to the question: in precisely what way predictions 
of the model~(\ref{EFT Lagrangian}) differ from those of the Starobinsky model.

It is very important that these quantum-gravitational  corrections, though being small compared to the bare $aR^2/2$ term,
can still be important both for dynamics of inflation and generation of perturbations, thus, they can be observable. The reason for
this is that the pure ${\cal L}=aR^2/2$ theory admits exact de Sitter solutions with {\em any} curvature. As a result, slow roll of 
curvature during inflation and the final graceful exit from it, as well as the slope of the scalar perturbation spectrum $n_s$ and 
the tensor-to-scalar ratio $r$, are governed by small corrections to this theory, namely, by the
Einstein term $\propto R$ in the Starobinsky $R+R^2$ model which is much less than the $R^2$ term during inflation. Thus, 
quantum corrections in the model ({\ref{EFT Lagrangian}) have to be compared to this small term, too. 

Of course, there is one more operator of canonical dimension four that can be added to~(\ref{EFT Lagrangian}), and that is 
$\Delta {\cal L}_{WW} = a_{WW}W_{\alpha\beta\gamma\delta}W^ {\alpha\beta\gamma\delta}$. However, to avoid 
problems with the ghost in the tensor sector~\cite{Fang:2012ca}, 
in this work we assume that $a$ in~(\ref{EFT Lagrangian}) is as anomalously
 large as required by observations (in fact, by the smallness of large-scale inhomogeneous perturbations in the Universe). 
On the other hand, there is no theoretical or observational reasons to assume such a large value for $a_{WW}$, so  
$a_{WW}\ll a$. As a result,  the $WW$ term can be neglected during the observable part of inflation.
(One may think that $R_{\mu\nu}R^{\mu\nu}$ should also be considered, but that term is expressible in terms
of the Gauss-Bonnet term which does not contribute to the equations of motion in 4 space-time dimensions,
$R^2$ and $W_{\alpha\beta\gamma\delta}W^ {\alpha\beta\gamma\delta}$, and hence need not be considered separately.)
Furthermore, there are higher dimensional operators of the form $a_n R^n/M_{\rm P}^{2(n-2)}$ ($n\geq 3$), {\it etc.},
but if none of $a_n$'s is anomalously large, their contribution will be unimportant during inflation, and hence can be neglected.
Indeed, a rough estimate can be made as follows. During the $R+R^2$ inflation, 
$R/M_{\rm P}^2\approx 2 N/(3a) \ll 1$, where $N$ is the number of e-foldings from the end of inflation ($1\ll N \lesssim 60$),
and hence the relative contribution of these terms compared to the non-leading Einstein term $M_P^2R/2$, as explained above, 
is  $\sim a_n(R/M_{\rm P}^2)^{n-1} \sim a_n (N/a)^{n-1)}\ll 1$ ($\forall n\geq 3$) as far as $Na_n^{1/(n-1)}\ll a$, see 
also~\cite{Huang14} in this connection. Thus, as the above argument suggests. these models do not suffer from large 
corrections coming from higher dimensional operators. However, with a suitable amount of fine tuning, it is possible to do 
away with lower dimensional operators, such that one still gets a viable inflationary model driven by higher dimensional 
operators of dimension four and higher~\cite{Marunovic:2016reh}.

We are not the first who consider an effective gravity inspired by asymptotic safety to drive inflation.
Notable initial attempts are due to Bonanno, Reuter and S
aueressig~\cite{Bonanno:2001xi,Bonanno:2001hi,Reuter:2005kb,Bonanno:2010bt} and more recently by Falls et al~\cite{Falls:2016wsa}.
Except in the most recent reference~\cite{Falls:2016wsa}, these works used a time-dependent cutoff which breaks general covariance
and therefore require a better justification.
More recently, Refs.~\cite{Weinberg:2009wa,Tye:2010an,Hindmarsh:2011hx} have renewed the idea that inflation may
be driven within an effective quantum gravity inspired by asymptotic safety, see also the recent review~\cite{Bonanno:2017pkg}. In contrast 
to our model, in these works inflation is driven by a (scale-dependent) cosmological constant, $\Lambda=\Lambda(\mu)$,
and the authors do not explain how inflation ends.
While our effective model takes into account the running of the coupling constants,
and in that respect it is  motivated by the recent results on asymptotically safe gravity theories~\cite{Demmel:2015oqa},
rigorously speaking it is not an asymptotically safe gravity model of inflation where inflation occurs close to the conformal point. 
Instead, in our model inflation occurs in the infrared regime rather far from the ultraviolet fixed point. 
Furthermore, the cosmological constant is in our model
assumed to be fine tuned to zero and inflation is driven by the $a(\mu,R)R^2$-term,
such that graceful exit problem is naturally solved.
A sufficiently long lasting inflation is obtained by assuming an anomalously large coefficient $a$, which is consistent with
the renormalization group equations, since $a$ appears as an integration constant~\cite{Demmel:2015oqa}.
Other notable papers that discuss inflation in effective theories inspired by quantum gravitational effects which, like us,
study inflationary models inspired by quantum corrections to the Starobinsky model,
include~\cite{Copeland:2013vva,CJSS15,Bamba:2014mua,BJTWZ14,Bonanno:2015fga}. 
Similar effective models can also arise from reconstruction 
of $f(R)$ gravity from observations~\cite{RCVZ14}.

This paper is organized as follows. In section~\ref{The model} we present our inflationary model.
After the background equations of motion are introduced
in~\ref{Background dynamics and properties of cosmological perturbations},
in~\ref{Cosmological perturbations in our model} we discuss the specifics of cosmological perturbations in our model
and in~\ref{Implementing constraints} we discuss how to implement the COBE constraint.
In section~\ref{Results} we present our main results, which include the dependence
of the scalar spectral index $n_s$, its running $\alpha$ and the tensor-to-scalar ratio $r$ on the parameter $b$
in  Eq.~(\ref{EFT Lagrangian}).
Finally, we conclude in section~\ref{Conclusion}. In the Appendix, an alternative derivation of $n_s$ and $r$ directly
in the Jordan frame is presented,

Here we adopt units in which the speed of light $c=1$ and the reduced Planck constant $\hbar\equiv h/(2\pi)=1$.


\section{The model}
\label{The model}

In a recent paper Demmel, Saueressig and Zanusso~\cite{Demmel:2015oqa} have considered quantum gravitational
corrections to Einstein general relativity and they found that at high energy scales quantum gravitational effects
generate a contribution to the effective action of the form $\Delta S_{\rm eff} =\int d^4 x  a(\mu)R^2/2$, where
$a$ is a parameter that slowly (logarithmically) varies with scale. By dimensional transmutation~\cite{Coleman:1973jx}
one can argue that, since $a(\mu)$ is dimensionless, it must be a function of a dimensionless quantity $R/\mu^2$, {\it i.e}
$a=a(R/\mu^2)=a_0[1 - b_0\ln(R/\mu^2)]$, where $R$ is the Ricci scalar.
While $a_0=a(R/\mu^2=1)$ is a free constant (to be fixed by measurements),
Ref.~\cite{Demmel:2015oqa} found that,
\begin{equation}
b_0 =\frac{40}{55296\pi^2a_0 + 95}
\,.
\label{b_0 Saueressig}
\end{equation}
In this work we assume
that, if a suitable renormalization group (RG) resummation is made, one gets an improved effective action of the form
\begin{equation}
S=\int d^4 x\sqrt{-g}\frac{1}{2}\left[M_{\rm P}^{2}R+\frac{a R^2}{1+b\ln(R/\mu^2)}+{\cal O}(R^3)\right]
\,,
\label{effective action}
\end{equation}
where $M_{\rm P}=(8\pi G)^{-1/2}$ is the reduced Plank mass, $a$ and $b$ are positive constants and $a\gg 1$.
For notational simplicity we have dropped the subscript $0$ on parameters $a$ and $b$ in~(\ref{effective action}). 
While $a$ is a free constant to be fixed by measurements, $b$ receives contributions
both from quantum effects of matter and gravitational fields that are calculable by perturbative methods within a given theory
and from threshold effects from the (unknown) Planck scale physics.
In this work we assume that the action~(\ref{effective action}) drives inflation.
According to the COBE normalization of scalar cosmological perturbations, 
the Hubble rate $H$ during the observable part of inflation,  
$H\sim H_{\rm I}\sim 10^{14}~{\rm GeV}$, is much smaller than the Planck energy,
$E_{\rm P}\simeq 1.2\times 10^{19}~{\rm GeV}$. For that reason
we expect that the higher dimensional operators ${\cal O}(R^3)$
in~(\ref{effective action}) present both a negligible contribution to evolution of the inflaton and to
measurable properties of cosmological perturbations in our model and, therefore, we neglect
these higher order terms  in the remainder of this work. While in pure gravity $a$ and $b$
are related by Eq.~(\ref{b_0 Saueressig}), adding matter may change that relation~\cite{Alkofer:2018fxj},
 and hence we shall relax that condition.
On the other hand, if  the action~(\ref{effective action}) is to be used as a model of inflation, then
the amplitude of scalar cosmological perturbations is fixed by the COBE normalization.
This fixes one relation between $a$ and $b$ in~(\ref{effective action}), such that ultimately our inflationary model
has one free parameter. We shall study how observable predictions of our model depend on that free parameter.

Note that in the limit of small $b$, the second term in the effective action~(\ref{effective action}) appears
to be equivalent to (a) $aR^2 - ab\ln(R/\mu^2)$ from Ref.~\cite{Demmel:2015oqa} as well as to (b)
$a\mu^4 (R/\mu^2)^{2-b}$. Inflation built based on the former form was also studied 
in~\cite{Gurovich:1979xg,BJTWZ14,RCVZ14}, while on the latter form -- in~\cite{CJSS15,CM15,Motohashi:2014tra,Liu:2018}.
One should point out however, that for any finite $b$ these models are {\it not} equivalent to~(\ref{effective action}).
Note also the crucial difference of our model from that considered in~\cite{RCVZ15} where the 
first (Einstein) term in the action~(\ref{effective action}) was absent. As explained above, this term, 
though being small compared to the second one during inflation, strongly affects slow roll of $R$ and 
the values of $n_s$ and $r$. In addition, it provides a graceful exit from inflation.  

Let us now proceed to analyze inflation governed by~(\ref{effective action}).
The action~(\ref{effective action}) is equivalent to,
\begin{equation}
S=\int d^4 x\sqrt{-g}\frac{1}{2}\left[f(\Phi)+\omega^2(R-\Phi)\right]
\,,
\label{effective action 2}
\end{equation}
where $f(\Phi)=M_{\rm P}^2\Phi+a \Phi^2/[1+b \ln(\Phi/\mu^2)]$,
$\Phi$ is a real scalar field (dubbed {\it scalaron} in~\cite{Starobinsky:1980te}) and $\omega=\omega(x)$
is a Lagrange multiplier (constraint) field (whose equation of motion imposes $\Phi=R$).
Now upon varying the action~(\ref{effective action}) with respect to $\Phi$ and solving the resulting equation, one obtains
\begin{equation}
f'(\Phi)-\omega^2=0,
\label{EOM for Phi}
\end{equation}
where
\begin{equation}
  f'(\Phi)= \frac{d f}{d \Phi}\equiv F(\Phi)
\,.
\label{F def}
\end{equation}
Inserting~(\ref{EOM for Phi}) into~(\ref{effective action})
results in an action  equivalent to~(\ref{effective action}),
\begin{equation}
S=\int d^4 x\sqrt{-g}\frac{1}{2}\left[f(\Phi)+F(\Phi)(R-\Phi)\right]
\,.
\label{effective action 3}
\end{equation}
Note that the scalaron $\Phi$ in~(\ref{effective action 3}) is non-minimally coupled to gravity {\it via}
the term $F(\Phi)R/2$. It is hence useful to refer to this form of the action as {\it Jordan frame}.
 It is well known that one can transform~(\ref{effective action 3}) to {\it Einstein frame} through a suitable
conformal transformation $g_{\mu\nu}=\Omega^2(x)g_{\mu\nu}^E$, where $\Omega=\Omega(x)$ is some 
still-to-be-specified local function.
By making use of the standard conformal transformation for the Ricci scalar and metric determinant $g$ one obtains,
\begin{eqnarray}
S&=&\int d^4 x\sqrt{-g_E}\frac{1}{2}\bigg[\Omega^{2}F(\Phi)
   \bigg(\!R_E-6g^{\mu\nu}_E \frac{\nabla_\mu^E \nabla_\nu^E\Omega}{\Omega}
\bigg)
  -\Omega^4\Big(F(\Phi)\Phi-f(\Phi)\Big)
\bigg].
\label{effective action 4}
\end{eqnarray}
By choosing the conformal function according to,
\begin{equation}
\Omega^{2}(\Phi)=\frac{M_{\rm P}^2}{F(\Phi)}
\,,
\label{Omega vs Phi}
\end{equation}
partially integrating the second term in the first line of Eq.~(\ref{effective action 4})
and dropping the resulting boundary term, the action of~(\ref{effective action 4}) becomes,
\begin{equation}
S=\int d^4 x\sqrt{-g_E}\left[\frac{M_{\rm P}^2}{2}R_E
 -3M_{\rm P}^2g^{\mu\nu}_E\frac{(\nabla_\mu^E\Omega)(\nabla_\nu^E\Omega)}{\Omega^2}
 -\frac12\Omega^4(F\Phi-f)\right]
\,,
\end{equation}
where $\Omega=\Omega(\Phi)$ through~(\ref{Omega vs Phi}).
We have got rid of the higher dimensional gravitational operator, but the prize is the emergence of a dynamical scalar field --
the scalaron field. Note that scalaron has a non-canonical kinetic term however and one can bring it to a canonical form
by the following transformation to {\it Einstein frame},
 \begin{equation}
\phi_E=-\frac{M_P}{2}\sqrt{6}\ln\big(\Omega^2(\Phi)\big)
\,,
\label{transformation to Einstein frame}
\end{equation}
where the field mapping is such that $\phi_E=0$ when $\Omega=1$.
Notice that instead of~(\ref{transformation to Einstein frame}) one could have chosen
a field transformation with the opposite sign.  In fact
that transformation is equivalent to~(\ref{transformation to Einstein frame}) in the sense that the resulting
Einstein frame potential would be the mirror image around $\phi_E=0$ of the potential obtained
by the transformation~(\ref{transformation to Einstein frame}).
When~(\ref{transformation to Einstein frame}) is exacted one obtains the following Einstein frame action,
\begin{equation}
S=\int d^4 x\sqrt{-g_E}\left[\frac{M_{\rm P}^2}{2}R_E-\frac{1}{2}g^{\mu\nu}_E(\partial_\mu\phi_E)(\partial_\nu\phi_E)
             -V_E(\phi_E)\right]
\,,
\label{effective action 5}
\end{equation}
where $V_E(\phi_E)$ denotes the Einstein frame potential,
\begin{equation}
V_E(\phi_E)=\frac{M^4_P}{2}\frac{F\Phi-f}{F^2}
\,,
\label{Einstein frame potential}
\end{equation}
and where, in light of Eqs.~(\ref{Omega vs Phi}), (\ref{transformation to Einstein frame}) and~(\ref{F def}),
\begin{equation}
 F(\phi_E) =M_{\rm P}^2\exp\bigg(\sqrt{\frac{2}{3}}\frac{\phi_E}{M_{\rm P}}\bigg)
= M_{\rm P}^2+\frac{a\Phi[2-b+2b\ln(\Phi/\mu^2)]}{[1+b\ln(\Phi/\mu^2)]^2}
\,.
\label{Einstein frame potential 2}
\end{equation}
This equation defines the frame transformation, $\phi_E=\phi_E(\Phi)$. Unfortunately, its inverse $\Phi=\Phi(\phi_E)$
is not a simple function that can be written in a closed form, so we write $V_E(\phi_E)$ as a function of $\Phi$,
but we keep in mind that $\Phi$ can be expressed in terms of $\phi_E$ via~(\ref{Einstein frame potential 2}).

 Taking the point of view that gravity is an effective field theory~\cite{Donoghue:1994dn},
the theory~(\ref{effective action 5}) can be (canonically) quantized. There are two dynamical fields in~(\ref{effective action 5}):
the inflaton $\phi_E$ and the graviton $g_{\mu\nu}$ that ought to be quantized. In what follows, we first discuss
the dynamics of classical fields (condensates) and then the (tree level) dynamics of quantum perturbations.

\subsection{Background dynamics and properties of cosmological perturbations}
\label{Background dynamics and properties of cosmological perturbations}

 The action~(\ref{effective action 5}) can be used to drive inflation, provided the quantum field $\hat \phi_E$
develops a large expectation value. If the field is approximately homogeneous with respect to a space-like hypersurface,
then it can be decomposed into its condensate (inflaton) and small perturbations as follows,
\begin{equation}
  \hat \phi_E(x) = \phi_{E0}(t) + \hat \varphi_E(x)
 \,,\qquad
\phi_{E0}(t) =\langle \hat \phi_E(x) \rangle \equiv {\rm Tr}\left[\hat\rho(t)\hat \phi_E(x) \right]
\,,
\label{inflaton: def}
\end{equation}
where $\hat\rho(t)$ denotes the density operator. Similarly, the metric tensor (in Einstein frame) can be written as,
\begin{equation}
  \hat g^E_{\mu\nu}(x) = g^{Eb}_{\mu\nu}(t)+  \delta \hat g^E_{\mu\nu}(x)
 \,,\qquad
 g^{Eb}_{\mu\nu}(t) =\langle \hat g^E_{\mu\nu}(x) \rangle = {\rm diag}\left(-1,a_E^2(t),a_E^2(t) , a_E^2(t)\right)
\,,
\label{inflaton: def}
\end{equation}
and $ \delta \hat g^E_{\mu\nu}(x)=a_E^2(t)  \hat h_{\mu\nu}(x)$, where $a_E=a_E(t)$ is the Einstein frame scale factor
and $\hat h_{\mu\nu}(x)$ is the (suitably rescaled) graviton perturbation in Einstein frame.
We work here in the traceless transverse (Lifshitz) gauge, in which $\hat h_{0\mu}=0$, $\partial_i \hat h_{ij}(x)=0=\hat h_{ii}(x)$.

\medskip

 The dynamics of the inflaton condensate is governed by the equation of motion,
\begin{equation}
 \ddot \phi_{E0}(t) + 3H\dot\phi_{E0}(t)+\frac{dV_E}{d\phi_{E0}}=0
\,,
\label{EOM phi E0}
\end{equation}
where we neglected any backreaction from quantum fluctuations. Analogously,
 evolution of the background geometry
is governed by the Friedmann (or FLRW) equations,
\begin{eqnarray}
 H_E^2 &\equiv& \left(\frac{\dot a_E}{a_E}\right)^2
    = \frac{1}{3M_{\rm P}^2}\left(\frac{\dot \phi_{E0}^2}{2}+V_E(\phi_{E0})\right)
\label{Friedmann 1}\\
 \dot H_E
    &=& -\frac{\dot \phi_{E0}^2}{2M_{\rm P}^2}
\,,
\label{Friedmann 2}
\end{eqnarray}
where $H_E$ is the Hubble parameter in Einstein frame and $\dot H_E = dH_E/dt$.

 Let us now consider scalar and tensor cosmological perturbations in the model~(\ref{effective action 5}).
It is convenient to decompose scalar and graviton perturbations in Fourier modes,
\begin{eqnarray}
\hat \varphi_E(t,\vec x) & =& \int \frac{d^3k}{(2\pi)^3}{\rm e}^{\imath \vec k\cdot \vec x}
    \left[\varphi(t,k)\hat a(\vec k)+\varphi^*(t,k)\hat a^\dagger(-\vec k)\right]
\nonumber\\
\hat h_{ij}(t,\vec x) & =& \sum_{\alpha = +,\times}\int \frac{d^3k}{(2\pi)^3}{\rm e}^{\imath \vec k\cdot \vec x}
    \left[\epsilon_{ij}^\alpha(\vec k)h(t,k)\hat b_\alpha(\vec k)+
           \epsilon_{ij}^\alpha(-\vec k)^*h^*(t,k)\hat b^\dagger_\alpha(-\vec k)\right]
\label{Fourier dec of perturbations}
\end{eqnarray}
where $k=\|\vec k\,\|$,  $\hat a(\vec k)$  and $\hat a^\dagger(\vec k)$ the annihilation and creation operators
for scalar perturbations, $\hat a(\vec k)|\Omega\rangle=0$ annihilates the vacuum state $|\Omega\rangle$,
 and  $\varphi(t,k)$ and $\varphi(t,k)^*$ are the two linearly independent solutions to the mode function
equation,
\begin{equation}
  \left(\frac{d^2}{dt^2} + 3H_E\frac{d}{dt} + \frac{k^2}{a_E^2}+\frac{d^2V_E}{d\phi_{E0}^2}\right)\varphi(t,k) = 0
\,.
\label{EOM mode functions}
\end{equation}
$ \epsilon_{ij}^\alpha(\vec k) $ in~(\ref{Fourier dec of perturbations}) are the spin 2 polarization tensors, obeying
 $\sum_\alpha  \epsilon_{ij}^\alpha(\vec k)\epsilon_{kl}^\alpha(-\vec k)^*
=(P_{ik}P_{jl}+P_{il}P_{jk}-P_{ij}P_{kl})/2$, $P_{ij}=\delta_{ij}-k_ik_j/k^2$,
 $\sum_{ij}  \epsilon_{ij}^\beta(\vec k)\epsilon_{ij}^\alpha(-\vec k)^*=\delta_{\alpha\beta}$,
  $\hat b_\alpha(\vec k)$  and $\hat b_\alpha^\dagger(\vec k)$ are the graviton annihilation and creation operators
($\hat b_{\alpha=+,\times}(\vec k)|\Omega\rangle=0$)
and $h(t,k)$ and $h^*(t,k)$ are the graviton mode functions that satisfy,
\begin{equation}
    \left(\frac{d^2}{dt^2} + 3H_E\frac{d}{dt} + \frac{k^2}{a_E^2}\right)h(t,k) = 0
\,.
\label{EOM mode functions: graviton}
\end{equation}
In  the zero curvature gauge (in which the spatial scalar graviton perturbation vanishes), scalar curvature perturbation is given by,
\begin{equation}
 {\cal R}(x) = -\frac{H_E}{\dot \phi_{E0}}\varphi(x)
\,,\qquad \dot\phi_{E0}^2 =  2\epsilon_E H_E^2 M_{\rm P}^2
\,.
\label{scalar curvature pert}
\end{equation}
Late time observers can measure properties of cosmological perturbations, which are characterized by the corresponding spectra,
which are defined as,
\begin{eqnarray}
   \langle \hat {\cal R}(t,\vec x)\hat {\cal R}(t,\vec x^{\,\prime})\rangle
            &=& \int \frac{dk}{k}{\rm e}^{\imath \vec k\cdot (\vec x-\vec x ^{\,\prime})} \Delta^2_R(t,k)
\label{scalar spectrum}\\
\langle \hat h_{ij}(t,\vec x)\hat h_{ij}(t,\vec x^{\,\prime})\rangle
            &=& \int \frac{dk}{k}{\rm e}^{\imath \vec k\cdot (\vec x-\vec x^ {\,\prime})} \Delta^2_t(t,k)
\label{graviton spectrum}
\end{eqnarray}
where during inflation the scalar and tensor spectrum can be calculated in terms of scalar and graviton mode functions as follows,
\begin{eqnarray}
 \Delta^2_R(t,k) &\equiv&  \Delta^2_s(t,k) = \frac{k^3}{8\pi^2\epsilon_EM_{\rm P}^2}|\varphi(t,k)|^2
\nonumber\\
 \Delta^2_t(t,k) & =& \frac{2k^3}{\pi^2M_{\rm P}^2}|h(t,k)|^2 = 16\epsilon_E \Delta^2_s
\,,
\label{spectra vs modes}
\end{eqnarray}
where we made use of Eqs.~(\ref{Fourier dec of perturbations}) and~(\ref {scalar curvature pert}).
When canonically normalized and to leading order in slow roll approximation,
the mode functions on super-Hubble scales can be approximated by,
$|\varphi|^2\simeq [H_*^2/(2k^3)][k/(aH)_*]^{n_s-1}$;
 an analogous approximation holds for the graviton mode functions, $|h|^2\simeq [H_*^2/(2k^3)][k/(aH)_*]^{n_t}$.

Astronomers usually parametrize the observed spectra as follows,
\begin{eqnarray}
  \Delta^2_s(k) &=&\Delta^2_{s*}\left(\frac{k}{k_*}\right)^{n_s(k)-1}
\nonumber\\
 \Delta^2_t(k) &=& \Delta^2_{t*}\left(\frac{k}{k_*}\right)^{n_t(k)}
\,,
\label{observer's spectra}
\end{eqnarray}
where $k_*$ is a fiducial comoving momentum usually chosen to be $k_*=0.05~({\rm Mpc})^{-1}$
or $k_*=0.002~({\rm Mpc})^{-1}$,
$\Delta^2_{s*}\equiv A_s$ and $\Delta^2_{t*}$ are the amplitude of scalar and tensor spectra evaluated at $k=k_*$
and $n_s$ and $n_t$ are the scalar and tensor spectral indices, respectively.
The COBE normalization~\cite{Ade:2015lrj} constrains
\begin{equation}
\ln(10^{10}A_s)=3.089\pm 0.036\quad  ({\rm at}\;\; k_*=0.05~{\rm MeV}^{-1})
\quad \Longrightarrow \quad A_s=( 2.20\pm 0.08)\times 10^{-9}
\,,
\label{COBE normalization}
\end{equation}
while from theory of scalar inflationary perturbations we know that
(to leading order in slow roll approximation),
\begin{equation}
A_s \simeq \frac{H_*^2}{16\pi^2\epsilon_{E^*}M_{\rm P}^2}
\,,
\label{COBE normalization 2}
\end{equation}
where $H_*$ and $\epsilon_{E*}$ are the Hubble parameter and
the principal slow roll parameter at the time when the perturbation with comoving momentum, $k=k_*$
crosses the Hubble radius (becomes super-Hubble) during inflation. Furthermore, observations constrain
the scalar spectral index $n_s$ and the tensor-to-scalar ratio, defined by
\begin{equation}
   r(k=k_*=0.05~{\rm Mpc}^{-1}) = \frac{\Delta^2_{t*}}{\Delta^2_{s*}}
\,.
\label{r definition}
\end{equation}
According to the Planck collaboration~\cite{Ade:2015lrj},
\begin{equation}
n_s =  0.9655 \pm 0.0062 \; (68 \%\; {\rm CL,\; Planck\; TT+lowP}, \alpha=0)
\,,
\label{observational limits: ns}
\end{equation}
when the running of the spectral index,
\begin{equation}
   \alpha\equiv \bigg[\frac{dn_s(k)}{d\ln(k)}\bigg]_{k=k_*}
\,,
\label{running alpha}
\end{equation}
is fixed to {\it zero}.  When the constraint on $\alpha$ is relaxed however, the error bars on $n_s$ increase somewhat
to become,
\begin{equation}
n_s  = 0.965\pm 0.010 \;\;\; (1\sigma\;{\rm error \;bars})
\quad {\rm and } \quad\alpha=-0.003\pm0.007
\,.
\label{observational limits: ns and alpha}
\end{equation}
When recent large scale structure (LSS) data are included~\cite{Palanque-Delabrouille:2015pga},
error bars on $n_s$ shrink and one finds preference for a negative running,
$n_s=0.963\pm 0.0045$ and $\alpha=-0.0104\pm 0.0031$. However, these results are still to be confirmed.
At this moment there are no measurements of tensor perturbations; instead the literature quotes upper bounds.
For example, the joint analysis BICEP2/Keck and Planck data found~\cite{Ade:2015tva}
$r<0.12~(95\%{\rm CL})$ and more recently~\cite{Array:2015xqh}
the BICEP2/Keck collaboration finds,
\begin{equation}
r<0.09\; (95\% {\rm CL},\; {\rm at}\; k_*=0.05~{\rm Mpc}^{-1})
\qquad ({\rm BICEP2/Keck})
\,.
\label{BICEP2/Keck constraint on r}
\end{equation}

\subsection{Cosmological perturbations in our model}
\label{Cosmological perturbations in our model}

In what follows we relate the observable parameters $n_s$, $r$ and $\alpha$
of scalar and tensor cosmological perturbations~(\ref{observer's spectra}) to our model in Einstein frame,
defined by Eqs.~(\ref{effective action 5}--\ref{Einstein frame potential 2}).
For a later use, we note that the Einstein frame potential~(\ref{Einstein frame potential}) can be written explicitly as,
\begin{equation}
V_E(\phi_E)
=\frac{M_{\rm P}^4}{2}
   \frac{a\Phi^2\big(1+b\ln\big[\frac{\Phi}{\mu^2}\big]\big)^2\big(1+b\ln\big[\frac{\Phi}{e\mu^2}\big]\big)}
   {\big\{M_{\rm P}^2\big(1+b\ln\big[\frac{\Phi}{\mu^2}\big]\big)^2
                             +2a\Phi\big(1+b\ln\big[\frac{\Phi}{\sqrt{e}\mu^2}\big]\big)\big\}^2}
\label{Einstein frame potential:3}
\end{equation}
where $\Phi=\Phi(\phi_E)$ is given in Eq.~(\ref{Einstein frame potential 2}) and $e\approx 2.81$. We shall also need
$d\Phi/d\phi_E$, which is easily obtained by differentiating~(\ref{Einstein frame potential 2}),
\begin{equation}
\frac{\partial \Phi}{\partial \phi_E}=\sqrt{\frac{2}{3}}\frac{\big(1+b\ln\big[\frac{\Phi}{\mu^2}\big]\big)
   \left(\big(1+b\ln\big[\frac{\Phi}{\mu^2}\big]\big)^2M_{\rm P}^2+a\Phi\big(2-b+2b\ln\big[\frac{\Phi}{\mu^2}\big]\big)\right)}
                    {a M_P\Big[(2-3b+2b^2)+b\ln\big[\frac{\Phi}{\mu^2}\big]\big(4-3b+2b\ln\big[\frac{\Phi}{\mu^2}\big]\big)\Big]}
\,.
\label{partial Phi partial phiE}
\end{equation}

 Most of inflationary models exhibit attractor behavior, which means that the physical parameters (such as the spectra) are expressible
in terms of the inflaton amplitude alone (in an attractor, $\dot\phi_E$ is a function $\phi_E$, and in the attractor known
as slow roll approximation, $\dot\phi_E$ and higher order time derivatives of the field are small in a well defined sense).
In what follows we apply the slow roll attractor results to our model~(\ref{effective action 5}--\ref{Einstein frame potential 2}).

From the canonical quantization of scalar and tensor
perturbations~(\ref{Fourier dec of perturbations}), (\ref{EOM mode functions}), (\ref{EOM mode functions: graviton})
and Eq.~(\ref{spectra vs modes})  one can show that when the inflaton is in
its attractor regime and when slow roll approximation applies,
$n_s$ and $n_t$ can be expressed in terms of geometric slow roll parameters
as follows,
\begin{equation}
  n_s = 1-2\epsilon_E - \eta_E
\,,\qquad
  n_t = -2\epsilon_E
\,,\qquad \epsilon_E = - \frac{\dot H_E}{H_E^2}
\,,\qquad \eta_E = \frac{\dot \epsilon_E}{H_E\epsilon_E}
\,.
\label{spectral index: geometric}
\end{equation}
In general, the spectral index $n_s$ is a function of $k$, and its running
$\alpha=dn_s/d\ln(k)$, can be expressed in terms of slow roll parameters as,
\begin{equation}
   \alpha =-\eta_E(2\epsilon_E+\xi_E)\,,\qquad \xi_E\equiv \frac{\dot\eta_E}{\eta_E H_E}
\,.
\label{alpha in terms of slow roll}
\end{equation}
In general $\alpha$ also depends on $k$. However, current observations are not precise enough for a detection of $\alpha$
and thus only upper limits on $|\alpha|$ are available. For that reason in this work by $\alpha$ we mean $\alpha(k_*)$.

One can express geometric slow roll parameters in terms of the more traditional slow roll parameters defined in terms of
derivatives of the inflaton potential $V_E$. For example, we have
\begin{eqnarray}
 \epsilon_V &=& \frac{M_{\rm P}^2}{2}\left(\frac{V_E^\prime}{V_E}\right)^2
,\qquad V_E^\prime \equiv \frac{dV_E}{d\phi_E}=\frac{d\Phi}{d\phi_E}\frac{dV_E}{d\Phi}
\nonumber\\
  \eta_V &=& M_{\rm P}^2 \frac{V_E^{\prime\prime}}{V_E}
\,,\qquad \quad\;\;\;\;\; V_E^{\prime\prime} = \left(\frac{d\Phi}{d\phi_E}\right)^2\frac{d^2V_E}{d\Phi^2}
                                                 + \frac{d^2\Phi}{d\phi_E^2}\frac{dV_E}{d\Phi}
\,,\nonumber\\
\nonumber\\
\xi_V^2 &=& M_{\rm P}^4\frac{V_E^\prime V_E^{\prime\prime\prime}}{V_E^2}
\,,\qquad
V_E^{\prime\prime\prime} = \left(\frac{d\Phi}{d\phi_E}\right)^3\frac{d^3V_E}{d\Phi^3}
                                               + 3\frac{d^2\Phi}{d\phi_E^2}\frac{d\Phi}{d\phi_E}\frac{d^2V_E}{d\Phi^2}
                                                 + \frac{d^3\Phi}{d\phi_E^3}\frac{dV_E}{d\Phi}
\,,
\label{traditional slow roll parameters}
\end{eqnarray}
where $V_E(\phi_E)$ and $d\Phi/d\phi_E$ are given in~(\ref{Einstein frame potential:3}--\ref{partial Phi partial phiE}).
Together with $d\ln(k)=d\ln(aH)$ (which expresses the fact that the amplitude of perturbations in slow roll get frozen
at super-Hubble scales) and  the Friedmann equations~(\ref{Friedmann 1}--\ref{Friedmann 2}),
Eqs.~(\ref{traditional slow roll parameters}) imply,
\begin{equation}
\epsilon_V=\epsilon_E
\,,\qquad
\eta_E = -4\epsilon_V-2\eta_V
\,,
\label{spectral index: relation}
\end{equation}
and hence
\begin{equation}
  n_s =1 -6\epsilon_V + 2\eta_V
\,,\qquad
  n_t= -2\epsilon_V
\,.
\label{spectral index: traditional 2}
\end{equation}
Furthermore, one can show that,
\begin{eqnarray}
  r &=& 16\epsilon_E = 16\epsilon_V = -8 n_t
\,\label{r in traditional slow roll}\\
  \alpha &=& 16\epsilon_V\eta_V-24\epsilon_V^2-2\xi_V^2
\,.
\label{alpha in terms of slow roll parameters}
\end{eqnarray}
Equation~(\ref{r in traditional slow roll}) is known as the one field consistency relation
 and it can be used {\it e.g.} to check whether inflation is driven by a single (inflaton) field.

Notice that, while $n_s$, $r$ and $n_t$~(\ref{spectral index: traditional 2}), (\ref{r in traditional slow roll})
 are of first order in slow roll parameters, the running of the spectral index
$\alpha$~(\ref{alpha in terms of slow roll parameters})
is of second order in slow roll parameters, and hence it is expected to be smaller than $n_s$, $r$ and $n_t$.
As we shall see below, this expectation is indeed borne out in our model.

Finally, a useful quantity to define is the number of e-foldings, which in Einstein frame
and in slow roll approximation can be calculated as follows,~\footnote{The formula~(\ref{number of efolds slow roll})
calculates the number of e-foldings in  the Einstein frame in the slow roll approximation. The more appropriate measure
of cosmological time is the number of e-foldings in Jordan frame,
since that is the original (physical) frame in which observations are made.
There is a simple relation between the number of e-foldings in the two frames. From Eq.~(\ref{Omega vs Phi}) and
$g_{\mu\nu}=\Omega^2 g_{\mu\nu}^E$ one sees that $\ln(a_J)=\ln(a_E)+\frac12\ln\left(M_{\rm P}^2/F(\Phi)\right)$,
such that the number of e-foldings in two frames are related as,
$N_J(t)=N_E(t)-\frac12\ln\left(F(\Phi)/M_{\rm P}^2\right)|_{\Phi_e}^{\Phi(t)}$, where ${\Phi_e}$ denotes the value of
$\Phi$ at the end of inflation. Numerical evaluation shows that the difference between $N_J$ and $N_E$
is at most a few percent, which means that results presented in terms of Einstein rather than Jordan frame
number of e-foldings will differ by at most a few percent.
Since the uncertainly in the number of e-foldings due to the unknown evolution of post-inflationary universe is
anyway at the level of ten percent, we can use the Einstein frame number of e-foldings
without introducing a significant new error. Nevertheless, the expression for the number of e-foldings in the Jordan 
frame $N_J$ is presented in the Appendix. 
}
\begin{equation}
 N\approx N_E = \int_{t}^{t_e} H_E(t^\prime) dt^\prime
            = \frac{1}{M_{\rm P}}\int_{\phi_{E}(t)}^{\phi_{Ee}}\frac{d\phi_E^\prime}{\sqrt{2\epsilon_E(\phi_E^\prime)}}
        = \frac{1}{M_{\rm P}}\int_{\Phi(t)}^{\Phi_e}
                       \frac{d\phi_E(\Phi^\prime)/d\Phi^\prime}{\sqrt{2\epsilon_V(\Phi^\prime)}}d\Phi^\prime
\,,
\label{number of efolds slow roll}
\end{equation}
where $t_e$ denotes the time at the end of inflation (at which $\epsilon_E=1$), and $\phi_{Ee}=\phi_E(t_e)$,
$\Phi_e=\Phi(t_e)$. Now equation~(\ref{number of efolds slow roll}) tells us how the number of e-foldings
depends on $\Phi$ (or equivalently $\phi_E$), while
Eqs.~(\ref{traditional slow roll parameters}--\ref{spectral index: relation}),
(\ref{spectral index: traditional 2}--\ref{alpha in terms of slow roll parameters}) relate $n_s$, $r$ and $\alpha$ to
$\Phi$ (or equivalently $\phi_E$). When taken together, and these two sets of (parametric) relations tell us
how $n_s$, $r$ and $\alpha$ depend on $N$. Unfortunately, the form of
the effective potential~(\ref{Einstein frame potential:3}) is rather complicated
such that we were unable to perform the integral~(\ref{number of efolds slow roll}). Nevertheless, with a help of
the symbolic package {\tt Mathematica}, we were able to plot the relevant curves.

To summarize, in this section we have shown how to parametrically express measurable quantities
 $n_s$, $r=-8n_t$ and $\alpha$ in terms of the number of e-foldings $N$. In the next sub-section we show how to
implement the existing observational constraints, which primarily pose a restriction on the number of e-foldings
and the amplitude of scalar cosmological perturbations.

\subsection{Implementing constraints}
\label{Implementing constraints}

 Apart from the (obvious) constraint on the scalar spectral index,
there are two principal constraints that we ought to impose on our inflationary model:
\begin{enumerate}
\item the amplitude of the scalar spectrum, also known as the COBE constraint~(\ref{COBE normalization 2});
\item the number of e-foldings $N$~(\ref{number of efolds slow roll}).
\end{enumerate}

 The COBE constraint is given in Eq.~(28), from which we can infer the value of the potential at
the moment when the fiducial comoving momentum $k_*$ crosses the Hubble scale during inflation, $k_*=(aH)_*$ as follows,
\begin{equation}
    \frac{V_E(\phi_{E*})}{M_{\rm P}^4}=48\pi^2\epsilon_{E*}A_s \simeq 3\times 10^{-10}\Big(\frac{r_{*}}{5\times 10^{-3}}\Big)
                                     \Big(\frac{A_s}{2.2\times 10^{-9}}\Big)
\,.
\label{COBE constraint}
\end{equation}

 The constraint on the number of e-foldings $N$ is not watertight, as it hangs on `reasonable assumptions'
on post-inflationary evolution. Assuming for example (a) that evolution of the Hubble parameter during inflation is given,
(b) that the scale of inflation is given and (c) that the Universe after inflation quite quickly (within one expansion time) reaches
radiation era scaling, then one can rather accurately estimate the number of e-foldings at which observable scales cross
the Hubble radius during inflation. However, there are no data that would unambiguously fix the scale of inflation,
or the precise evolution of the Hubble parameter during or after inflation. To incorporate this uncertainty, usually
one plots physical parameters for several values of $N$. In this work we take the reasonable range of $N$ to be
$N\in [50,65]$.

In the analysis of our inflationary model presented in section~\ref{Cosmological perturbations in our model}
we enforce the COBE constraint as shown in~(\ref{COBE constraint}),
and we show results for $N=50$ and $N=65$. Our model~(\ref{effective action}) contains two free parameters $a$ and $b$
which determine
the scale of inflation $H_{E*}$ at $N\in[50,65]$. Imposing the COBE constraint fixes one relation between $a$ and $b$,
leaving one free parameter. Our results are shown as a function of that free parameter, which for definiteness we choose to be $b$,
while $a=a(b)$.

\begin{figure}[h!]
 \centering
   \includegraphics[width=0.8\textwidth]{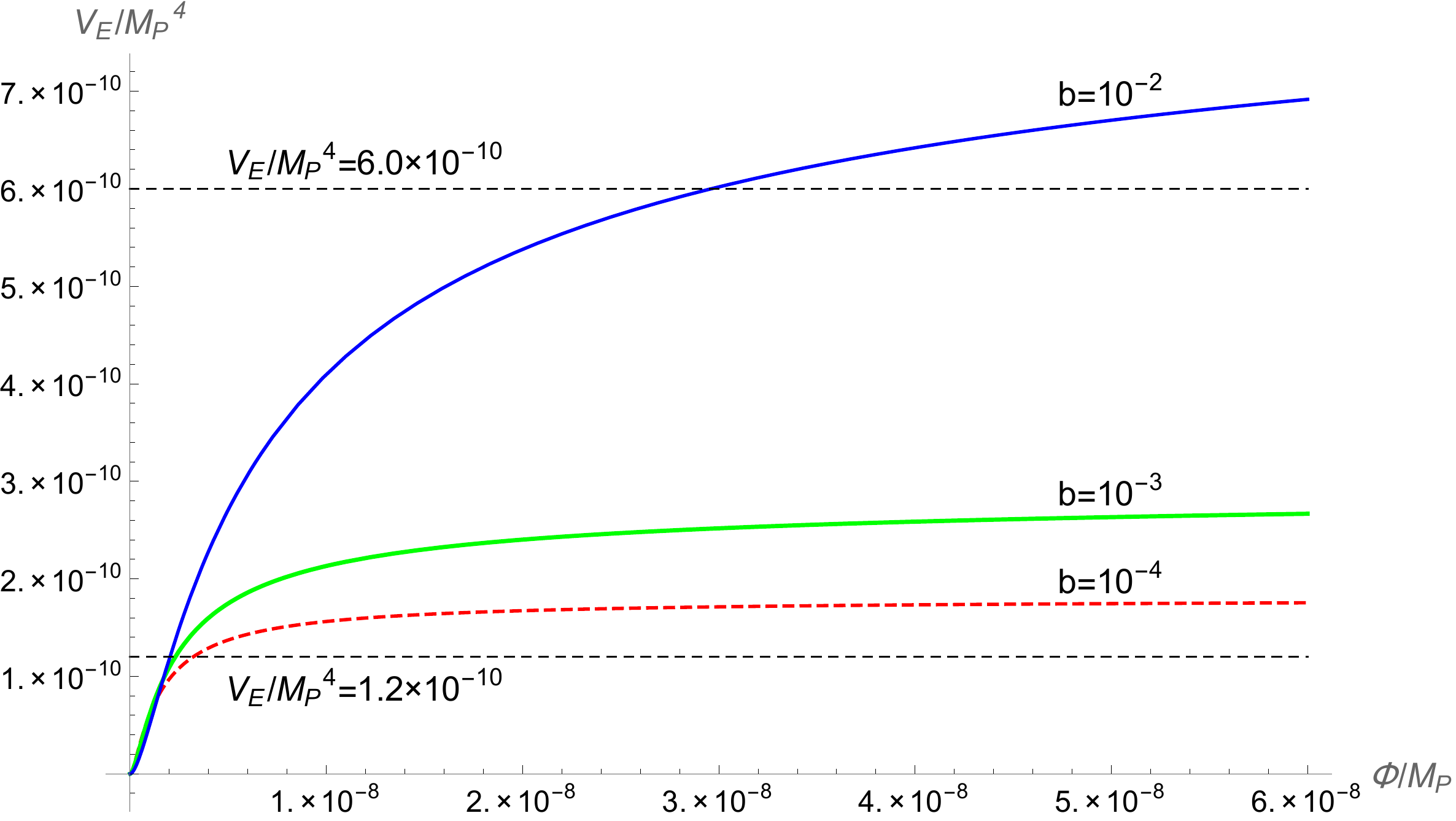}
 \caption{\footnotesize The Einstein frame potential $V_E$ from Eq.~(\ref{Einstein frame potential:3})
 is shown for $b=10^{-2}$  (top blue curve), $b=10^{-3}$ (middle green curve) and $b= 10^{-4}$ (lower red dashed curve).
In this figure $a=7\times 10^8$, $\mu=10^{-5}H_E$.
 The upper (lower) black dashed curve shows $V_E=6\times 10^{-10}M_{\rm P}^4$ ($V_E=1.2\times 10^{-10}M_{\rm P}^4$)
and they represent the upper and lower limits of the effective potential corresponding to the COBE constraint~(\ref{COBE constraint})
evaluated for $r=10^{-2}$ ($r=2\times 10^{-3}$).}
\label{figure 1}
 \end{figure}
 To get an impression on how our potential looks in Einstein frame, in figure~\ref{figure 1}
 we show the potential $V_E=V_E(\Phi)$ defined in Eq.~(\ref{Einstein frame potential:3}), where
(at scale $\mu=10^{-5}\simeq H_{E*}$)
the values of the couplings
are, $a=7\times 10^8$ and $b=10^{-2}$ (top curve), $b=10^{-3}$ (central curve) and $b=10^{-4}$ (bottom curve).
For convenience we shall also use $\tilde a$ defined by,
$a=10^{10}\tilde a$.
As $b$ decreases and for large values of $\Phi$ the potential becomes flatter and flatter. In the limit when
$b\rightarrow 0$ (Starobinsky inflation) and as $\Phi\rightarrow \infty$, the potential becomes exactly flat.

 \begin{figure}[h!]
 \centering
  \includegraphics[width=0.8\textwidth]{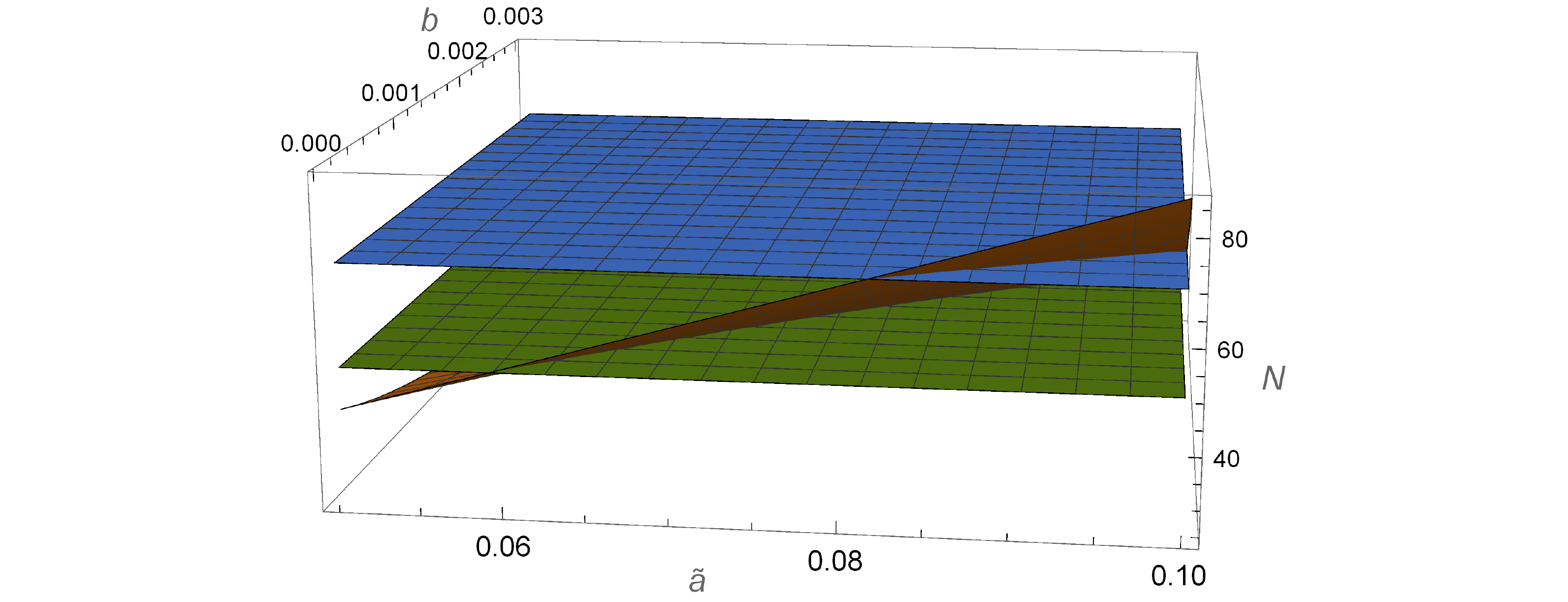}
 \caption{\footnotesize The number of e-foldings $N$ as a function of $\tilde a=10^{-10}a$ and $b$.
 The blue and green planes represent $N=50$ and $N=65$, respectively.}
\label{figure 3}
 \end{figure}
Figure~\ref{figure 3} shows the allowed range of parameters $\tilde a=10^{-10}a$ and $b$
for which one gets $N$ in the range $[50,65]$
(the COBE constraint~(\ref{COBE constraint}) is imposed).

 \begin{figure}[h!]
 \centering
  \includegraphics[width=0.6\textwidth]{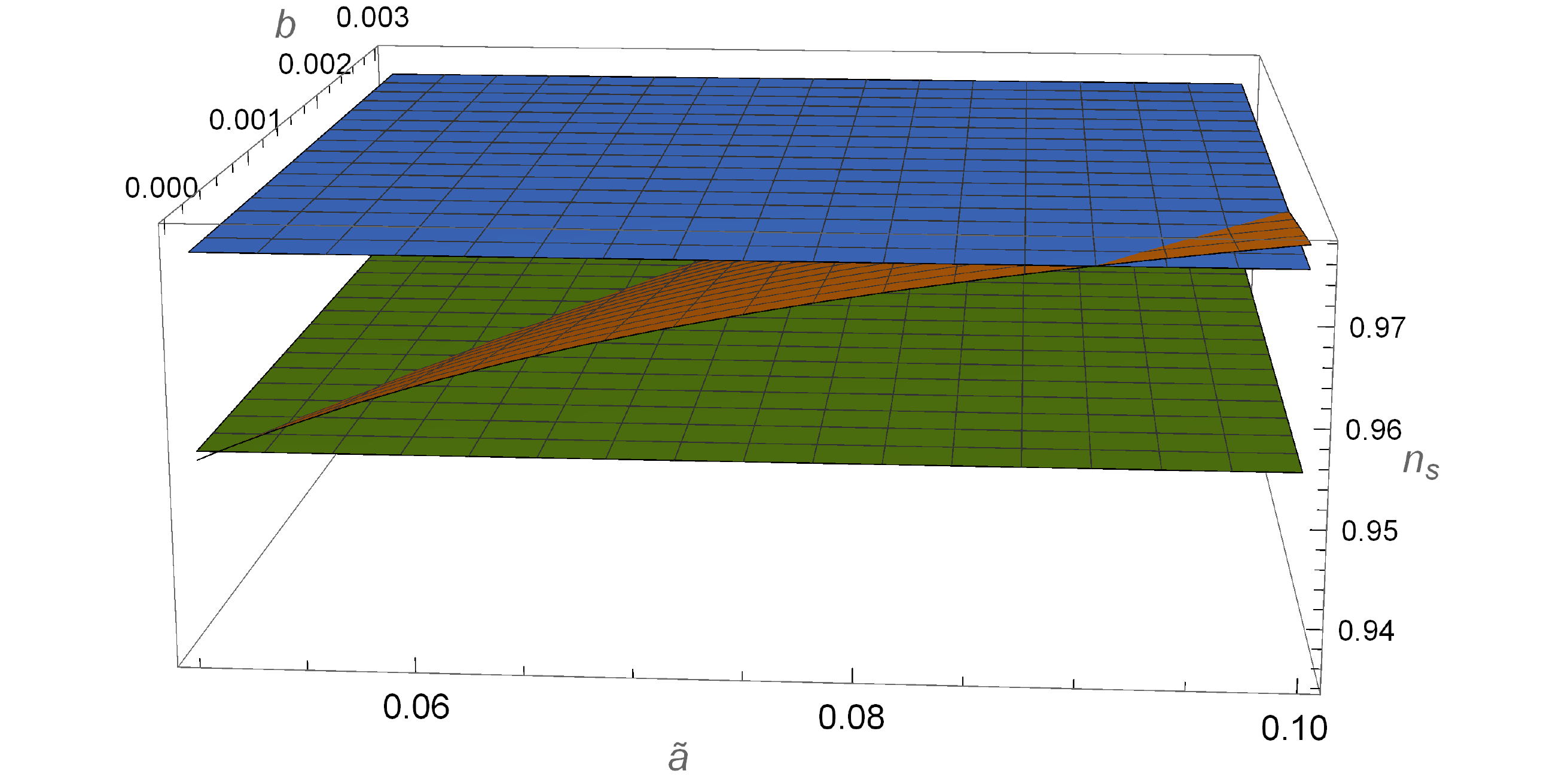}
 \caption{\footnotesize The scalar spectral index $n_s$ as a function of parameters $\tilde a=10^{-10}a$  and $b$.
The allowed range of $n_s$ is between the blue and green planes.}
\label{figure 4}
 \end{figure}
Next, in figure~\ref{figure 4} we show typical values of $\tilde a$ and $b$ for which our model yields the scalar spectral index
$n_s$ consistent with observations (the COBE constraint~(\ref{COBE constraint}) is imposed).

From these results it follows that, when $b\ge 10^{-3}$, our inflationary model predictions
depend significantly on $b$. However when $b\ll 10^{-3}$,
model predictions are to a large extent independent on $b$ and they can be well approximated by those of the Starobinsky model.


\section{Results}
\label{Results}

In this section we present our main results,  {\it i.e.} we plot the scalar spectral index $n_s$~(\ref{spectral index: traditional 2}),
its running $\alpha=dn_s/d\ln(k)$~(\ref{alpha in terms of slow roll parameters})
and tensor-to-scalar ratio $r$~(\ref{r in traditional slow roll})
evaluated at the fiducial comoving momentum $k=k_*=0.05~{\rm Mpc}^{-1}$
(the tensor spectral index $n_t$ is in our model given by the consistency relation,  $n_t=-r/8$ and it is
therefore not an independent observable). Even though it appears that $n_s$, $\alpha$ and $r$ are given in terms
of the three slow roll parameters $\epsilon_E$, $\eta_E$ and $\xi_E$, they are not all independent. In fact,
our model~(\ref{effective action}) is specified by two parameters, $a$ and $b$ (defined at a fiducial scale $\mu$),
and therefore in slow roll approximation and when the COBE normalization is imposed,
one free parameter remains (the number of e-foldings is fixed by the
value of the field, $\Phi=\Phi(\phi_E)$). That means that, for any given $N$, the prediction of our model
can be represented by a two-dimensional plane on the three dimensional configuration space spanned by $(n_s,r,\alpha)$.
The constraint on $n_s$ is then the three dimensional region on that configuration space defined by
$n_{s,{\rm min}}\simeq 0.955\leq n_s\leq n_{s,\rm max}\simeq 0.975$, see Eq.~(\ref{observational limits: ns and alpha}).
Rather than showing this three dimensional space,
for the sake of clarity we show in what follows its three cross sections with the planes $(n_s,r)$,
$(n_s,\alpha)$ and $(r,\alpha)$.

\begin{figure}[h!]
 \centering
  \includegraphics[width=0.6\textwidth]{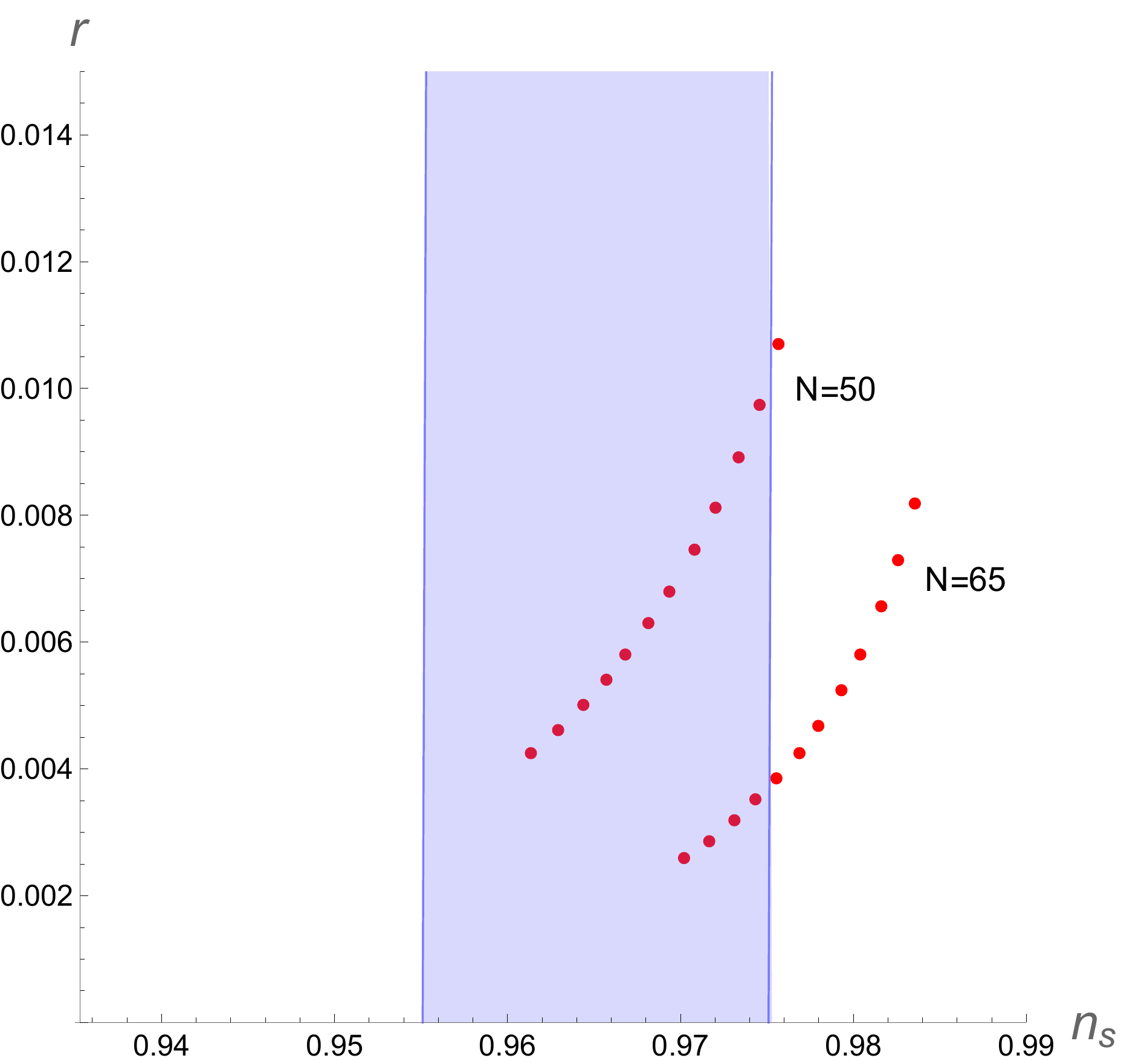}
 \caption{The tensor-to-scalar ratio $r$ as a function of the scalar spectral index $n_s$ for $b$ in the range from $10^{-4}$ to  $10^{-2}$. We show two curves: $N=50$ (upper red dotted curve) and  $N=65$ (lower red dotted curve).
The shadowed region shows the allowed values of $n_s$~(\ref{observational limits: ns and alpha}).
The current upper limit on
the tensor-to-scalar ratio, $r<0.09$~(\ref{BICEP2/Keck constraint on r})  is outside the plot's range.
The COBE constraint~(\ref{COBE constraint}) is imposed and $\mu=10^{-5}H_E$.}
\label{figure 5}
 \end{figure}
In figure~\ref{figure 5} we show how the tensor-to-scalar ratio $r$ depends on the scalar spectral index $n_s$.
The upper (lower) red dotted curve shows $r=r(n_s)$ for $N=50$ ($N=65$), $b$ ranges from $10^{-4}$ to $10^{-2}$,
and $a$ is fixed by the COBE constraint. For very small values of $b$ one recovers the predictions of the $R^2$-model,
$r\in [2,4]\times 10^{-3}$, and $n_s$ lies in the sweet spot of the allowed $n_s$.
When $b$ increases however, both $r$ and $n_s$ increase, such that when $r$ exceeds $r\simeq 10^{-2}$,
$n_s$ becomes too large to be consistent with observations.
Notice that even a modest increase in $r$ can have very beneficial consequences for detectability of tensor modes.
Namely, while advanced future satellite CMB probes (such as COrE~\cite{Bouchet:2011ck}
and LiteBIRD~\cite{Matsumura:2013aja})
can detect $r\sim 3\times 10^{-3}$ with an accuracy of few standard deviations, if
$r\simeq 1\times 10^{-2}$ these probes can claim discovery ({\it i.e.} more than $5 \sigma$ detection), and moreover
even Stage IV of earth-based CMB observatories can detect $r\simeq 10^{-2}$.

\begin{figure}[h!]
 \centering
  \includegraphics[width=0.6\textwidth]{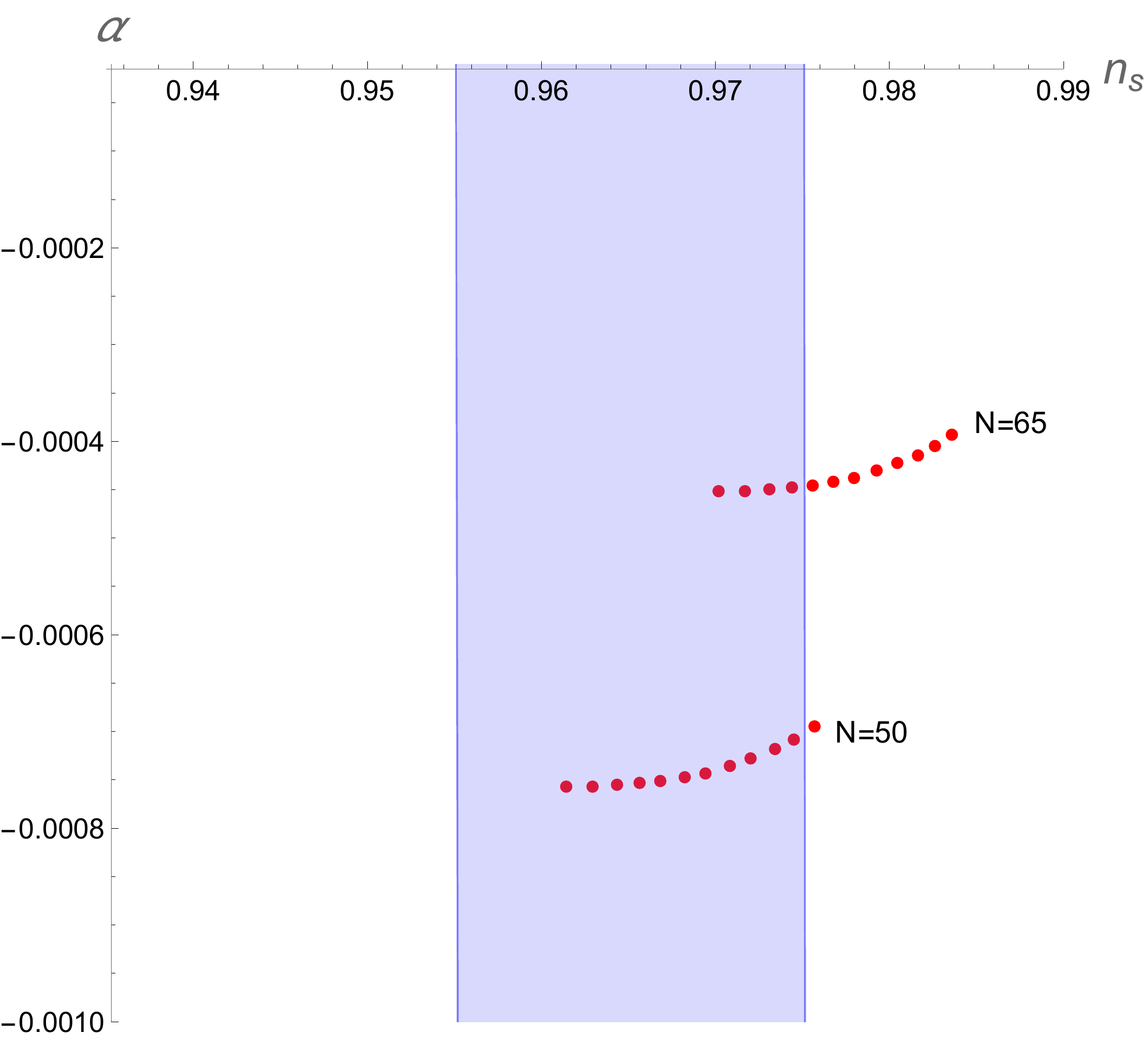}
 \caption{The scalar spectrum index running parameter $\alpha$ as a function of the scalar spectral index $n_s$ for $b$
in the range from $10^{-4}$ to  $10^{-2}$. We show two curves: $N=50$ (upper dotted curve) and  $N=65$ (lower dotted curve).
The shadowed region shows the allowed values of $n_s$ in Eq.~(\ref{observational limits: ns and alpha}).
The current Planck Collaboration limits on
the running of scalar spectral index, $\alpha\in[-0.010,0.004]$ lie outside the plot's range.
The COBE constraint~(\ref{COBE constraint}) is imposed and $\mu=10^{-5}H_E$.}
\label{figure 6}
 \end{figure}
In figure~\ref{figure 6}, in which we impose identical constraints as in figure~\ref{figure 5}, we show
$\alpha=\alpha(n_s)$ for $N=65$ (upper red dots) and for $N=50$ (lower red dots) with $b\in [10^{-4},10^{-2}]$.
As in figure~\ref{figure 5}, as $b$ increases, $n_s$ increases, eventually leaving the allowed range for $n_s$ when
$b>10^{-2}$. Note that  $\alpha$ does not depend significantly on $b$, so the logarithmic dependence $\ln(R/\mu^2)$
in the effective action~(\ref{effective action}) does not affect detectability of $\alpha$.
In order to detect $\alpha$ in our model, we need an improvement of about one order of magnitude with respect
to current observations, which is
possible only with space based observations (such as COrE), but unlikely with earth-based measurements.
Nevertheless, together with a detection of $r$, a detection of $\alpha$ would represent an important milestone
in testing the model proposed in this work.

\begin{figure}[h!]
 \centering
  \includegraphics[width=0.6\textwidth]{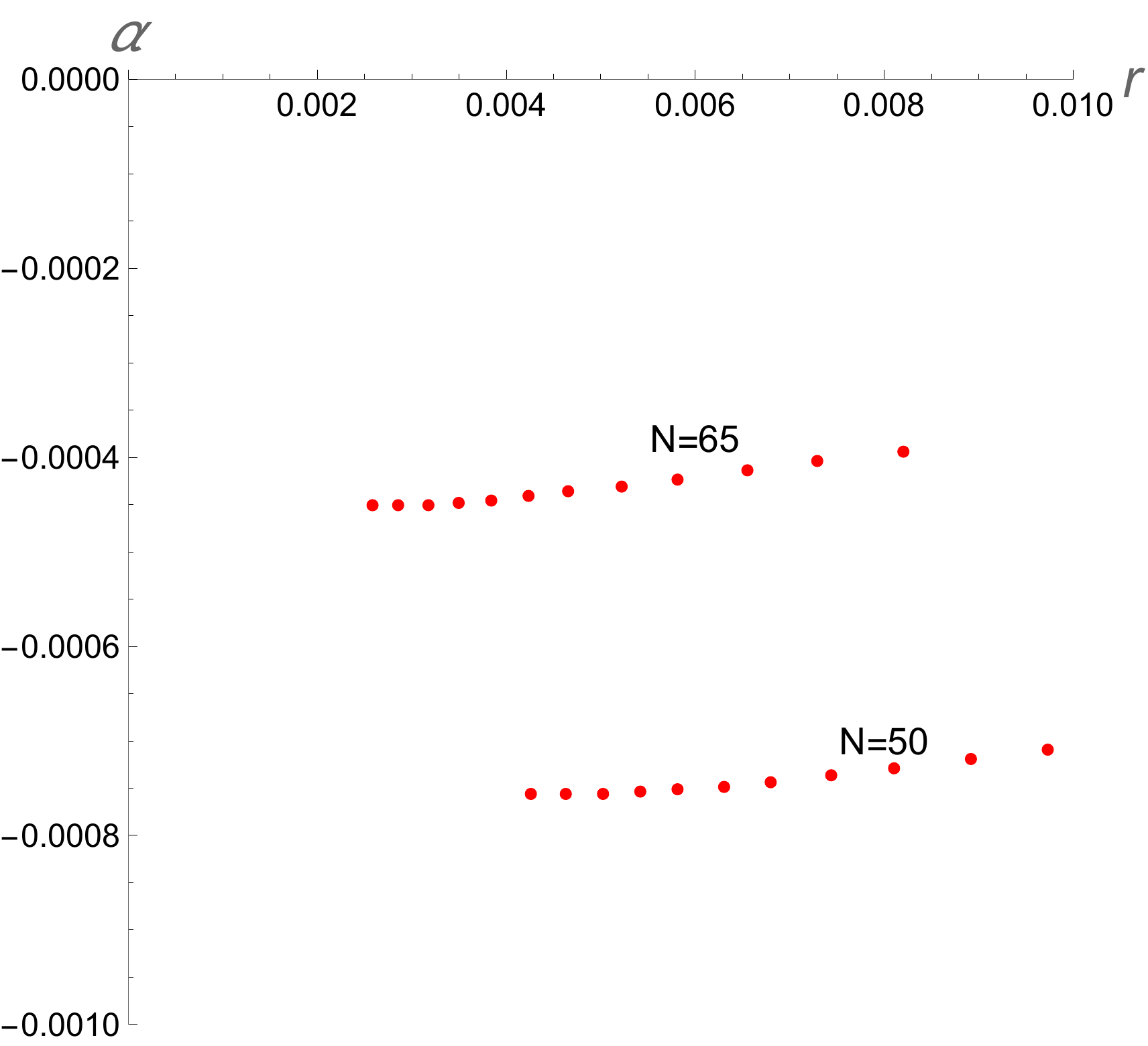}
 \caption{The  the running of scalar spectrum index $\alpha$ as a function of the tensor-to-scalar ratio $r$ for $b$
in the range from $10^{-4}$ to  $10^{-2}$. We show two curves: $N=50$ (upper dotted curve) and  $N=65$ (lower dotted curve).
The shadowed region shows the allowed values of $n_s$ in Eq.~(\ref{observational limits: ns and alpha}).
The current upper limit on
the tensor-to-scalar ratio, $r<0.09$~(\ref{BICEP2/Keck constraint on r})  is outside the plot's range.
The COBE constraint~(\ref{COBE constraint}) is imposed and $\mu=10^{-5}H_E$.}
\label{figure 7}
 \end{figure}
For completeness,
in figure~\ref{figure 7} we show $\alpha=\alpha(r)$ for $N=65$ (upper curve) and $N=50$ (lower curve).
As expected, as $b$ increases from $10^{-4}$ to $10^{-2}$, $r$ increases
(reaching $r\sim 10^{-2}$ when $b\sim 10^{-2}$) and $\alpha$ does not change much.


\section{Conclusion}
\label{Conclusion}

 We consider an effective gravity model~(\ref{effective action}) motivated by Weinberg's proposal
that gravity might be renormalizable in a weaker sense, known as
 {\it asymptotic safety}~\cite{Weinberg:1976xy,Hawking:1979ig,Weinberg:2009bg,Weinberg:2016kyd}.
 This model is supported both by perturbative studies of 
 quantum gravity~~\cite{Stanyukovich:1965,DeWitt:1967yk,DeWitt:1967uc,Ginzburg:1971,Gurovich:1979xg} as well as by more recent
studies~\cite{Demmel:2015oqa} based on functional renormalization group approach to quantum gravity.
Our model can be considered as an improvement of Starobinsky $R^2$ inflation. Our main
results can be stated as follows:
\begin{enumerate}
\item[$\bullet$] The effective theory of gravity~(\ref{effective action}) represents a viable model of inflation, provided $a$ and
$b$ are appropriately chosen, {\it i.e.} $a\sim 10^8-10^{9}$ (as implied by the COBE constraint) and $b\leq 10^{-2}$ (in order
that the scalar spectral index $n_s$ agrees with observations);

\item[$\bullet$] When $b=0$ we recover the predictions of the Starobinsky $R^2$ inflationary model, which is currently
a viable model of inflation and for which
the tensor-to-scalar ratio is about, $r\simeq 3\times 10^{-3}$;

\item[$\bullet$] As $b$ increases from $b=0$ to about $b\simeq 10^{-2}$, $r$ increases from
about $3\times 10^{-3}$ to about $1\times 10^{-2}$, which is much easier to detect by future
CMB polarization observatories, such as LiteBIRD and COrE. This increase in $r$ represents
 the main result of this work. Namely, observing $r$ in the range $r\in [5,10]\times 10^{-3}$
(and significantly larger than the value in the Starobinsky model, $r=3\times 10^{-3}$),
$n_s\simeq 0.970-0.975$ (which is about one standard deviation higher than the current central  value for $n_s$)
and $\alpha\simeq -5\times 10^{-4}$ (and significantly different from zero) would represent observational
evidence that would lend support to gravity as an asymptotically safe theory
and that it exhibits an ultraviolet fixed point on super-Planckian energy
scales. We are not aware of any other observations that could test gravity on (super-)Planckian scales.
Moreover, measuring $r\sim 10^{-2}$ would allow future space missions to test the consistency relation, $r=-8n_t$,
which represents an important consistency test of one field inflationary models. 
Owing to the smallness of $r$, the consistency relation
will be much harder to test in the original Starobinsky model.

\end{enumerate}

At first sight our model might seem unnatural, because the required value of $a$ is very large.
However, from the observational point of view it is the observed smallness of inhomogeneous perturbations
in the Universe  that determines this large value unambiguously. Also, from the theoretical point of view, 
one should keep in mind however, that the value of $a$ in Eq.~(\ref{effective action})
is not fixed by quantum loop effects. Indeed, adding an arbitrary finite coefficient $\Delta a$ to $a$
in~(\ref{effective action})  is consistent with quantum loop effects, {\it i.e.} one can fix $a=a(\mu)$
arbitrarily (in principle by observations) at some reference scale $\mu$, and then $a(\mu,R)$ will run
according to a suitable renormalization group equation, for more details see~\cite{Demmel:2015oqa}.
On the other hand, once one fixes $a$, one cannot arbitrarily choose $b$; the value of $b$ depends on
the bare Lagrangian, {\it i.e.} on the particle content of the theory, and for pure gravity  the relation
is given by~(\ref{b_0 Saueressig}), in which case $b\sim 10^{-4}$.
It would be of interest to investigate precisely how the relation between
$b$ and $a$ depends on the field content of the theory, and for what type of theories one gets
the values of $n_s$, $r$ and $\alpha$ that are of interest for observational cosmologists.
We intend to investigate that question in future work.
As a final remark, we note that measuring $b$ would provide us with 
some information (albeit very limited) on the structure of perturbative quantum gravity at the 
grand unified scale (of inflation) and (via the threshold effects) 
even on the physics at and beyond the Planck scale. 


\acknowledgments

A.S. is supported by the RSF grant 16-12-10401. L.L. is funded by the Chinese Scholarschip Council (CSC). 
T.P. and L.L. acknowledge supprot from the D-ITP consortium, 
a program of the NWO that is funded by the Dutch Ministry of 
Education, Culture and Science (OCW). This work is part of the research programme of the Foundation for Fundamental Research on Matter (FOM), which is part of the Netherlands Organisation for Scientific Research (NWO).


\section{Appendix}

Here we present the derivation of expressions for the power spectra of scalar and tensor perturbations generated during inflation in modified $f(R)$ gravity with
\begin{equation}
S=\frac{1}{2}\int \, d^4x\sqrt{-g}\, f(R)
\,,
\label {fR}
\end{equation}
directly in the physical (Jordan) frame, such that $f(R)=M_P^2R$ in GR.  
It becomes simple if one uses the fact~\cite{ABS10} that slow-roll inflation  in this class of models occurs for the range of $R$ where $f(R)$ is close to $R^2$, more exactly, $f(R)=A(R)R^2$ where $A(R)$ is a slowly changing function of $R$ :
\begin{equation}
|A'(R)|\ll \frac{A(R)}{R}~,~~|A''(R)|\ll \frac{A(R)}{R^2}
\,,
\label{A}
\end{equation}
where a prime denotes the derivative with respect to $R$. The conditions (\ref{A}) are the analogues of the slow-roll conditions for the Einstein frame potential $V_E(\phi_E)$ flatness.
They may be satisfied either over some interval of $R$ values, or even in one point $R=R_0$ only.

Then, from the trace equation of $f(R)$ gravity (written in the presence of matter with the energy density $\rho_m$ and the pressure 
$p_m$ for generality),
\begin{equation}
\frac{3}{a^3}\frac{d}{dt}\left(a^3\frac{df'(R)}{dt}\right) -Rf'(R)+2f(R) = 
\rho_m - 3p_m~,
\label{trace}
\end{equation}
an expression follows for the number of e-folds in Jordan frame during inflation counted from its end ($R=R_{\rm end}$) back in time
\begin{equation}
N(R) =-\frac{3}{2}\int_{R_{\rm end}}^{R} d\tilde R \, \frac{A(\tilde R)}{A'(\tilde R)\tilde R^2}\gg 1~.
\label{N-Jordan}
\end{equation}
Note that the condition $A'<0$ is needed for the correct evolution during inflation and the graceful exit from it to the region of small curvature.

The power spectrum of tensor perturbations (summed over polarizations) can be directly obtained from the corresponding expression for inflation in GR first derived in~\cite{Starobinsky:1979ty} by the substitution
$M_{\rm P}^2\to M_{\rm P,\, eff}^2=df/dR\approx 2AR$:
\begin{equation}
\Delta_t^2(k)=\frac{1}{12A(R_k)\pi^2}
\,,
\label{tensor}
\end{equation}
where, as usually, the index $k$ means that the corresponding quantity is estimated at the moment $t=t_k$ when each spatial Fourier mode of perturbations crosses the Hubble radius during inflation:  $k=a(t_k)H(t_k)$. Note, however, that $df/dR$ has to be calculated with better accuracy in order to find the correct value of the slope $n_t$ of the tensor spectrum.

Thus, at present $N(k)=\ln(k_{\rm end}/k)$ where $k_{\rm end}$ is the comoving wave vector of perturbations which crossed the Hubble radius at the end of inflation. $k_{\rm end}/a_0$ (where $a_0$ is the present value of the scale factor $a(t)$) is a few orders of magnitude smaller than
the present CMB temperature $T_{\gamma}$. Its exact value depends, in particular, on duration of the epoch of the scalaron decay and creation and heating of matter after inflation.

The easiest way to find the scalar perturbation spectrum is to use the $\delta N$ formalism first introduced in~\cite{S82} (and  even in the fully non-linear regime) according to which
spatial inhomogeneity of total number of e-folds during inflation leads to the following value of the scalar perturbation ${\cal R}$:
\begin{equation}
{\cal R}({\bf r})=\delta N_{tot}({\bf r})=\frac{dN(R)}{dR}\delta R({\bf r})
\,,
\label{deltaN}
\end{equation}
where $\delta R$ is estimated at the characteristic time of its Hubble radius crossing during inflation. This method was applied, in particular, in~\cite{S83} where the quantitatively correct expressions for the power spectra of scalar and tensor perturbations in the Starobinsky model were first obtained, see also~\cite{KM86} and numerous later papers.\footnote{The correct estimate of the slope of the scalar power spectrum $n_s$ in this model was found even earlier in~\cite{MCh81}.} Using (\ref{N-Jordan}) and the correctly normalized rms value of $\delta R$ fluctuations, we get
\begin{equation}
\Delta_{\cal R}^2(k)=\frac{A(R_k)}{64\pi^2A'(R_k)^2R_k^2} = -\left(96\pi^2\,\frac{dA(R(N))}{dN}\right)^{-1}~.
\label{scalar}
\end{equation}
Thus, 
\begin{equation}
r\equiv \frac{\Delta_t^2(k)}{\Delta_{\cal R}^2(k)}=
\frac{16A'(R_k)^2R_k^2}{3A(R_k)^2}
\label{r}
\end{equation}
and it can be checked that $r=-8n_t$. For general inflationary models in $f(R)$ gravity, the power spectrum of scalar perturbations was derived in~\cite{HN96}.

For the Starobinsky model, $A(R)=M_p^2\left(\frac{1}{R}+\frac{1}{6M^2}\right)$ where $M$ is the scalaron mass after the end of inflation (in particular, in flat space-time). The first term in $A$ is much less than the second one during inflation, so it may be neglected in $A$ itself, but not in its derivative $A'$. So, 
\begin{equation}
R(N)= 4NM^2,~~\Delta_{\cal R}^2(k)=\frac{M^2N^2(k)}{24\pi^2M_P^2},~~
n_s=1-\frac{2}{N(k)}, ~~r=\frac{12}{N^2(k)}~.
\label{Smodel}
\end{equation}

In our model, 
\begin{equation}
A=\frac{M_P^2}{R}+\frac{a}{1+b\ln(R/\mu^2)}
\label{ourA}
\end{equation}
with $a\gg 1$. Using Eq. (\ref{N-Jordan}), we get with sufficient accuracy:
\begin{equation}
N(R)=\frac{3a}{2}\int_{R_{\rm end}}^R d\tilde R \frac{1+b\ln\frac{\tilde R}{\mu^2}}
{M_P^2[1+b\ln(\tilde R/\mu^2)]^2+ab\tilde R}~.
\label{Nour}
\end{equation}
Let us assume that the logarithmic correction is small, $b\ll1$, $b\ln({R}/{\mu^2})\ll 1$.
Then  Eq. (\ref{Nour}) can be integrated analytically and
\begin{equation}
1+\frac{abR}{M_P^2}=\exp\left(\frac{2bN}{3}\right)~.
\label{small-b}
\end{equation}
If $bN\ll 1$, then $abR \ll M_P^2$ and we return to the results (\ref{Smodel}) for the Starobinsky model
with $a=M_P^2/6M^2$. In the opposite limit $bN\gg 1,~abR \gg M_P^2$ (but still $R\ll M_P^2$ that requires $ab\gg 1$), 
\begin{equation}
R=\frac{M_P^2}{ab}\exp\left(\frac{2bN}{3}\right),~~\Delta_{\cal R}^2(k)=\frac{1}{64\pi^2ab^2}~,~~n_s\approx 1, ~~r=\frac{16ab^2}{3}~.
\label{new-regime}
\end{equation}
Thus, it may not be reached for the observable range of perturbation wavelengths due to too large values of $n_s$ 
and $\Delta_{\cal R}^2r\approx 1/(12 \pi^2)$ which are independent both of $k$ and of the model parameters $a,b$. 
These analytical results help to understand the numerical results of section III.

\vskip -0.2cm

\eject

\bibliography{mybibfile}

\end{document}